\documentclass[12pt,preprint]{aastex}

\shorttitle{The eclipsing blue straggler V228 in 47~Tuc}
\shortauthors{Kaluzny et al.}

\begin{document}

\title{The Clusters AgeS Experiment (CASE). II. \\
The Eclipsing Blue Straggler OGLEGC-228 
in the Globular Cluster 47~Tuc\footnote{This paper utilizes data 
obtained with the 6.5-meter Magellan Telescopes located at Las Campanas 
Observatory, Chile.}
}

\author{J.~Kaluzny\altaffilmark{2}, 
I.~B.~Thompson\altaffilmark{3},
S.~M.~Rucinski\altaffilmark{4}, 
W.~Pych\altaffilmark{2}, 
G.~Stachowski\altaffilmark{2},
W.~Krzeminski\altaffilmark{5},
G.~S.~Burley\altaffilmark{3}}

\altaffiltext{2}{Copernicus Astronomical Center, Bartycka 18,
00-716 Warsaw, Poland; (jka,pych,gss)@camk.edu.pl}

\altaffiltext{3}{Carnegie Observatories, 813 Santa Barbara St.,
Pasadena, CA 91101-1292; (ian,burley)@ociw.edu}

\altaffiltext{4}{David Dunlap Observatory, Department of Astronomy and 
Astrophysics, University of Toronto, P.O. Box 360, Richmond Hill, 
ON L4C 4Y6, Canada; rucinski@astro.utoronto.ca}

\altaffiltext{5}{Las Campanas Observatory, Casilla 601, La Serena, 
Chile; wojtek@lco.cl}

\begin{abstract}
We use photometric and spectroscopic observations of the eclipsing
binary OGLEGC-228 (V228)  to derive the masses, radii, and luminosities
of the component stars. Based on measured systemic velocity, 
proper motion and distance, the system is a blue straggler 
member of the globular cluster 47~Tuc.
Our analysis shows that V228 is a semi-detached Algol. 
We obtain $M=1.512\pm 0.022\,M_\odot$, 
$R=1.357\pm 0.019\,R_\odot$, $L=7.02\pm 0.050\,L_\odot$
for the hotter and more luminous primary component and
$M=0.200\pm 0.007\,M_\odot$, $R=1.238\pm 0.013\,R_\odot$,
$L=1.57\pm 0.09\,L_\odot$ for the Roche lobe filling secondary.
\end{abstract}


\section{INTRODUCTION}
\label{intro}

Blue straggler (BS) stars are defined by their location
above and to the blue of the
main-sequence turn-off on the color-magnitude diagram of
their parent population. Since their discovery in the 
globular cluster M3 \citep{sandage} BS have been identified
in many  globular and open clusters \citep{piotto,demarchi}, as well
as in the field \citep{carney}. The currently most popular BS
formation mechanisms are  mass transfer in a 
close binary \citep{mccrea} and  merger of 
two stars induced by a close encounter \citep{benz}. 
There are numerous examples of confirmed
binary BS in open clusters and for a few  masses of both 
components have been accurately established from
analyses of radial and light curves 
(see for example \citet{sandquist2003}). In the case of globular clusters,
the sample of candidate BS includes several contact binaries
\citep{rucinski2000} as well as a sizable population of 
SX Phe pulsating variables which have been detected in over a dozen  clusters
\citep{rodriguez2000}.  
However, until now not a single, direct mass determination is
available for a binary BS belonging to a globular cluster.
\citet{demarco05} used $HST$ spectra to estimate masses for 24 
apparent single BS from 3 clusters (see also \citet{shara}). 
They derived an average mass of 1.07~$M_{\odot}$  with a 
wide range of uncertainty  for the masses of individual objects.
There thus seems to be evidence for both BS formation mechanisms
in open and globular clusters.
 
The eclipsing binary OGLEGC-228 (hereinafter V228) was discovered
by \citet{kal98} during a survey for variable stars 
in the field of the globular cluster 47~Tuc. 
They presented a $V$ band light curve for the variable
and found an orbital period of $P=1.1504$~d. 
Further $VI$ photometry of V228 along with a finding chart 
was published by \citet{weldrake} (star WSB V7 in their catalog).
On the color-magnitude diagram of the cluster, the variable
occupies a position near the top of the blue straggler sequence.

In this paper we report results of photometric and spectroscopic
observations aimed at a determination of absolute parameters of V228.
The data were obtained as a part of a long-term CASE project 
conducted at Las Campanas Observatory \citep{kal05}. Sections~\ref{phot}
and \ref{period} describe the photometry of the variable and an
analysis of its orbital period. Section~\ref{spec} presents the
radial velocity observations. The combined photometric and 
spectroscopic element solutions are given in Section~\ref{comb}
while the membership to 47~Tuc is discussed in Section~\ref{memb}.
The last Section~\ref{disc} discusses the properties of V228
in the 47~Tuc context.

\section{PHOTOMETRIC OBSERVATIONS AND REDUCTIONS}
\label{phot}

The photometric data were obtained with the 1.0-m Swope 
telescope at the Las Campanas Observatory using the SITE3 CCD camera 
at a scale of 0.435$\arcsec$/pixel.  
Most of the images were taken with the detector subrastered 
to $2048\times 3150$ pixels, but occasionally we also used 
subrasters of $2048\times 2150$ or $2048\times 700$ pixels.
Most of observations were collected during the 1997, 1998 and 2001 
observing seasons. The same set of $BV$ filters was used for all
observations. Some additional data
were obtained with the $V$ filter in 2003 and 2004 with a goal of 
refining the period of the system.
Exposure times ranged from 120~s to 200~s
for the $V$ filter and from 180~s to 240~s for the $B$ filter.
The raw data were pre-processed with the IRAF-CCDPROC
package\footnote{IRAF is distributed by the National Optical Astronomy
Observatories, which are operated by the Association of Universities
for Research in Astronomy, Inc., under cooperative agreement with the
NSF.}.  The time series photometry was
extracted using the ISIS-2.1 image subtraction package \citep{alard,alard2}.
To minimize the effects of a variable point spread function
we used $600\times 600$ pixel sub-images in the analysis with the variable
located in the center of the field.
Transformation of instrumental photometry to the standard
$BV$ system was accomplished using measurements of 198
standard stars from Stetson's catalog \citep{stet00} which are present 
in the observed field.   

\begin{figure}
\figurenum{1}
\label{fig1}
\plotone{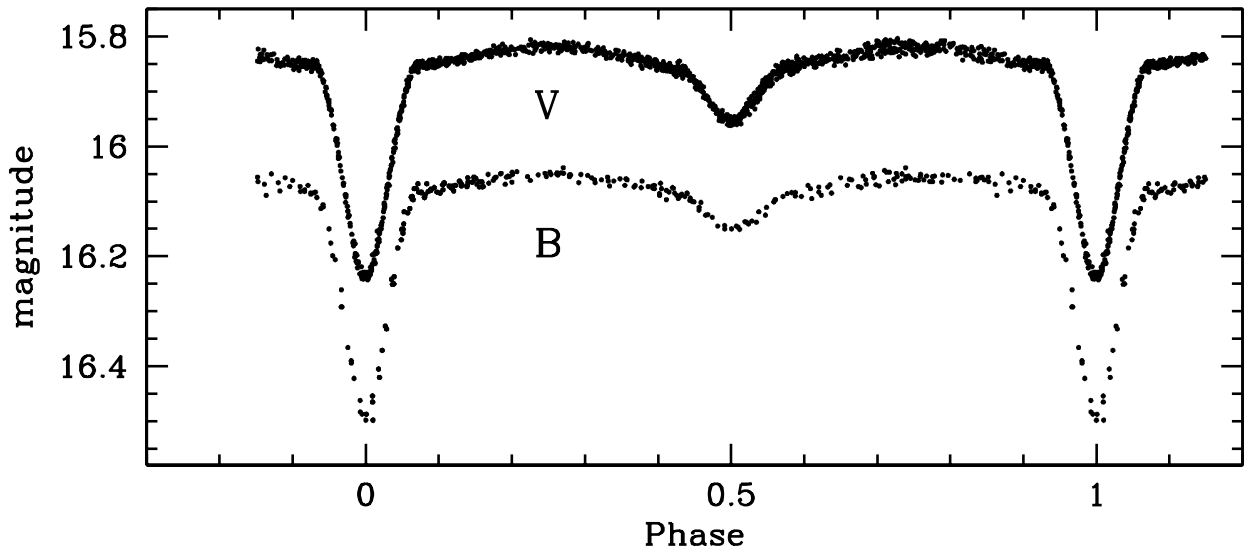}
\caption{
The phased $BV$ light curves of V228.
}
\end{figure}

In Fig.~\ref{fig1} we show the $BV$ light curves of V228 phased with 
the ephemeris given in the next subsection. These curves contain a total 
of 1199 and 274 data points for $V$ and $B$, respectively.
The colors and magnitudes of V228 at minima and at quadratures are listed 
in Table~\ref{tab1}. The quoted errors do not include possible systematic
errors of the zero points of the photometry which we estimate at about
0.01~mag.

\section{THE ORBITAL PERIOD}
\label{period}

>From the available data we measured 9 times of
eclipses for V228; their values, along with errors determined 
using the method of \citet{kwee} are given in Table~\ref{tab2}.
The first listed minimum is based on the OGLE-I data from \citet{kal98}. 
The $O-C$ values listed in the table correspond to the linear ephemeris:
\begin{equation}
Min I = HJD~2,451,064.82019(16) + 1.15068618(14) 
\end{equation}
determined from a least squares fit to the data. A linear
ephemeris provides a good fit and there is no evidence for any
detectable period change during the interval 1993--2004 covered by the
observations.

\section{SPECTROSCOPIC OBSERVATIONS}
\label{spec}

Spectroscopic observations of V228 were obtained
with the MIKE Echelle spectrograph \citep{bern03} on the Magellan~II
(Clay) telescope of the Las Campanas Observatory.
The data were collected during 5 observing runs 
between 2003 August 16 and 2004 October 4. For the current analysis we used 
the data obtained with the blue channel of MIKE
covering the range from 380 to 500 nm
with a resolving power of $\lambda / \Delta \lambda \approx 38,000$.
All of the observations were obtained with a $0.7\times 5.0$ arcsec slit
and with $2\times 2$ pixel binning. At 4380 \AA\ the resolution was
$\sim$2.7 pixels at a scale of 0.043~\AA/pixel.
The seeing ranged from 0.7 to 1.0~arcsec. 

The spectra were first processed using a pipeline developed by Dan
Kelson following the formalism of \citet{kel03,kel07} and then analyzed
further using standard tasks of the IRAF/Echelle package.
Each of the final individual spectra consisted of two 600--900~s exposures
interlaced with an exposure of a thorium-argon lamp. We obtained 40
spectra of V228. 
For the wavelengths interval  $400-500$ nm the average signal-to-noise 
ratios ranged between 14 and 39, depending on the observing conditions.
In addition to observations of the variable, we also
obtained high S/N  spectra of radial velocity template stars.

\begin{figure}
\figurenum{2}
\label{fig2}
\plotone{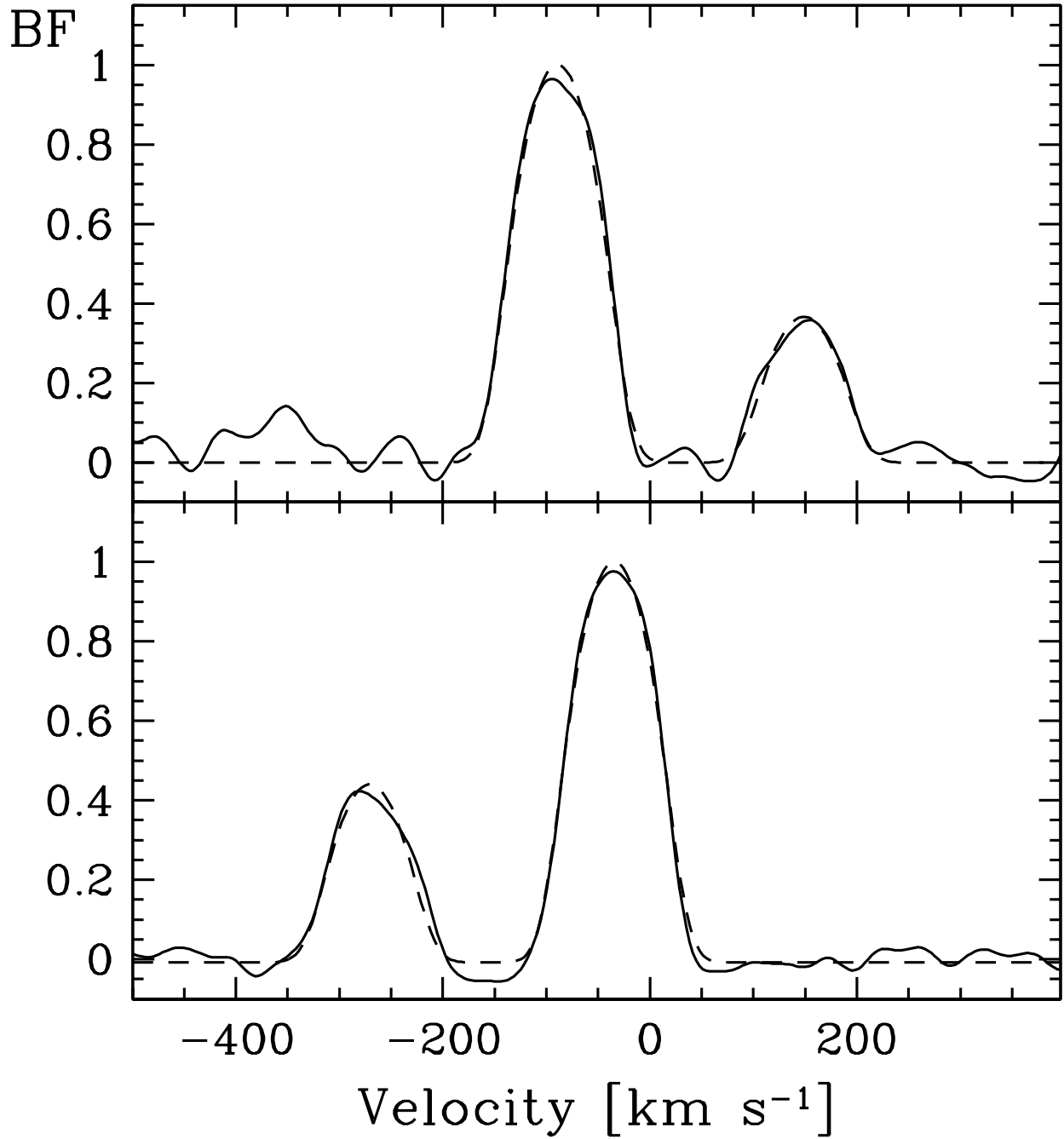}
\caption{
Broadening Functions extracted from the spectra of V228 obtained 
near the first (upper panel) and second (lower panel) quadratures. 
The dashed lines show fits of a model BF to the observed ones.
Note that the two components appear to be well detached from each 
other. However, our solution presented 
in Sections~\ref{comb} -- \ref{disc} shows that
the secondary fills it Roche lobe which is relatively small because of
the small mass ratio, $q=0.13$.}
\end{figure}

We have analyzed the spectra of V228 for radial velocity variations
using a code based on the broadening function (BF) formalism 
of \citet{ddo7}. A spectrum of HD~138549 was used as a template.
According to \citet{nordstrom}, the relevant properties of 
HD~138549 are: $V_{rad}=11.6$~km/s, $V \sin i= 1$~km/s, 
${\rm [Fe/H]}=+0.01$ and  $T_{eff}=5457$~K. 
We used the spectra in the wavelength range from 400 nm to 495 nm
excluding the Balmer series lines. 
Figure~\ref{fig2} presents examples of fitting a model
to the BFs calculated for two spectra taken near opposite quadratures.  
Our radial velocity
measurements of  V228 are listed in Table~\ref{tab3},
with last two columns giving 
residuals from the spectroscopic orbit solution, as
presented in the next section.  
The current implementation of the BF method does not give reliable
estimates of internal errors of the measured radial velocities; 
in the following analysis the velocities for a given
component were weighted according to the rms of
its $O-C$ residuals from the fitted spectroscopic orbit.

\section{COMBINED ANALYSIS OF THE LIGHT AND RADIAL VELOCITY CURVES}
\label{comb}

We analyzed the light  and radial velocity curves of V228  using
the Wilson-Devinney model \citep{wd71} as implemented in the
PHOEBE package \citep{prsa05}. We adopted an iterative  
scheme in which the light and radial velocity curves were fitted 
independently and alternately.

The following parameters were adjusted in the light curve solution:
the orbital inclination $i$, the gravitational potentials $\Omega_{1}$ and 
$\Omega_{2}$, the effective temperature of the secondary $T_{2}$ and the
relative luminosities $L_1$/$L_2$ in $B$ and $V$. 
The mass ratio was fixed at the value resulting from the spectroscopic 
solution. The temperature of 
the primary, $T_1$ was determined from the dereddened color index 
$(B-V)_{1}$ using the calibration of \citet{worthey}. We adopted
an interstellar reddening of $E(B-V)=0.04$ following \citet{harris}
and the metallicity $\rm{[Fe/H]}=-0.67$ from \citet{alves-brito}.
In the first iteration, the color index of the primary was assumed to
be the same as the observed color index of the binary at the quadrature,
leading to an initial $T_1=7630$~K. In the following iterations we used 
the previous step solution to disentangle the
magnitudes and colors of both components at the maximum light.
For the primary component, the bolometric albedo and the 
gravitational-brightening coefficients were set at values appropriate 
for stars with radiative envelopes: $A_{1}=1.0$ and $g_{1}=1.0$ while
for the cooler secondary we used  values appropriate
for stars with convective envelopes: $A_{2}=0.5$ and $g_{2}=0.32$.
An attempt was also made to solve  the light curves 
with $A_{2}=1.0$ and $g_{2}=1.0$, but the obtained fit was 
substantially poorer than the one derived  with $A_{2}=0.5$ 
and $g_{2}=0.32$. Specifically, the $\chi^{2}$ quantity increased 
by a factor of 2.6 and a systematic dependence of residuals on phase
became apparent in both minima.   
In the photometric solution the mass ratio was
set to the value measured from the spectroscopic solution.
The free parameters in the spectroscopic solution were the semi-major
axis $a$, the systemic velocity $V_0$ and the mass ratio $q=M_2/M_1$.
The solutions were started with a detached configuration, but 
converged quickly to a semi-detached configuration with 
the secondary component filling its Roche lobe. 
The starting value of the mass ratio was established from the
ratio of the spectroscopic radial velocity semi-amplitudes $K_2$ and
$K_1$, as derived from the preliminary sine curve fits. 

\begin{figure}
\figurenum{3}
\label{fig3}
\plotone{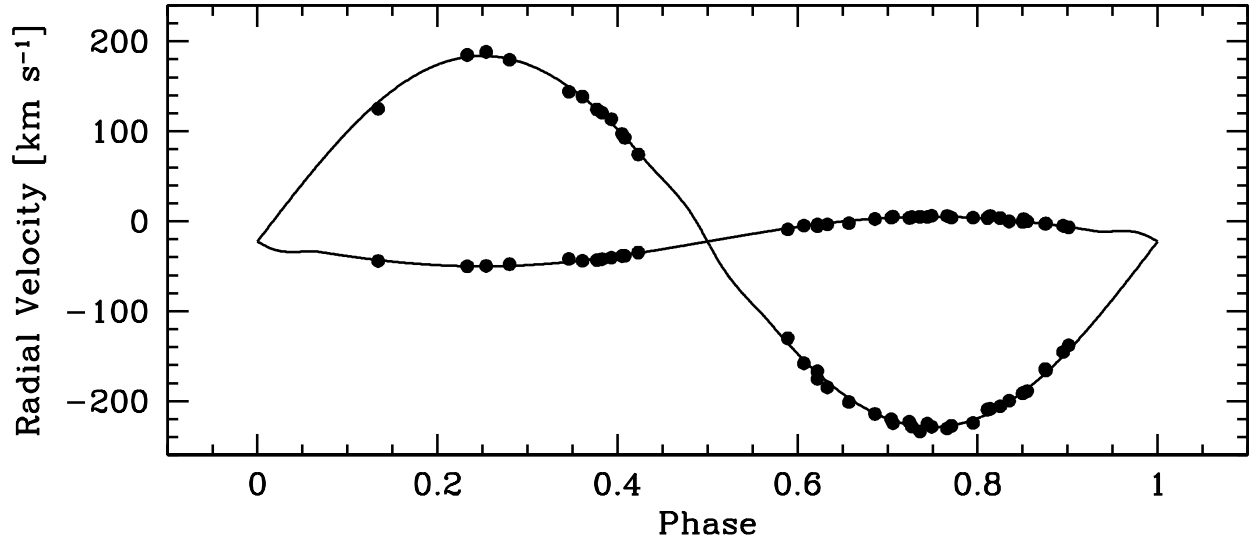}
\caption{
The spectroscopic observations and adopted radial velocity
orbit for V228.
}
\end{figure}

\begin{figure}
\figurenum{4} 
\label{fig4} 
\plotone{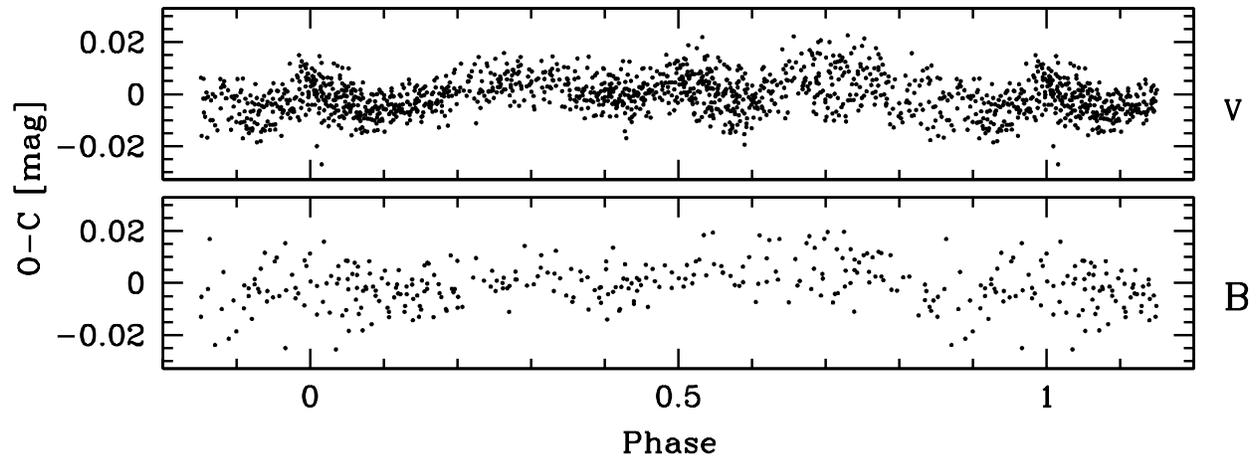}
\caption{
The residuals from the light curve solution in the $BV$ bands.
}
\end{figure}

The derived parameters of the spectroscopic orbit are
listed in Table~\ref{tab4}. 
Figure~\ref{fig3} shows the computed radial velocity curves 
together with radial velocity measurements. 
The light curve solution is listed in Table~\ref{tab5} and 
the residuals between the observed and synthetic light curves are
shown in Figure~\ref{fig4}. The quantities listed in the 
last column of Table~\ref{tab5} are the
weighted averages of the values obtained from the solutions for the 
$V$ and $B$ filters. One may notice that the parameters derived from 
solutions based on $V$ and $B$ photometry are consistent with each other.

Using the luminosity ratios from the light curve solution
one may obtain the observed visual magnitudes
of  the components of V228 at the maximum light. We obtained
$V_1=16.064 \pm 0.002$ , $B_1= 16.219 \pm 0.002$, $V_2=17.741 \pm 0.010$
and $B_2=18.391\pm 0.006$ where the errors represent the
respective uncertainties in the solution and do not include the
contribution from zero point uncertainties of our photometry of about 
0.01~mag.
For a reddening of $E(B-V)=0.04$, the de-reddened color index of the 
secondary component is  $(B-V)_{20}=0.610\pm 0.018$ which
implies an effective temperature of $T_{2}=5685\pm 85$~K 
according to the calibration of \citet{worthey}. It is encouraging to see
that $T_2$ derived this way is consistent with the value resulting from
the light curve solution listed in Table~\ref{tab5}. 

\begin{figure}
\figurenum{5} 
\label{fig5} 
\plotone{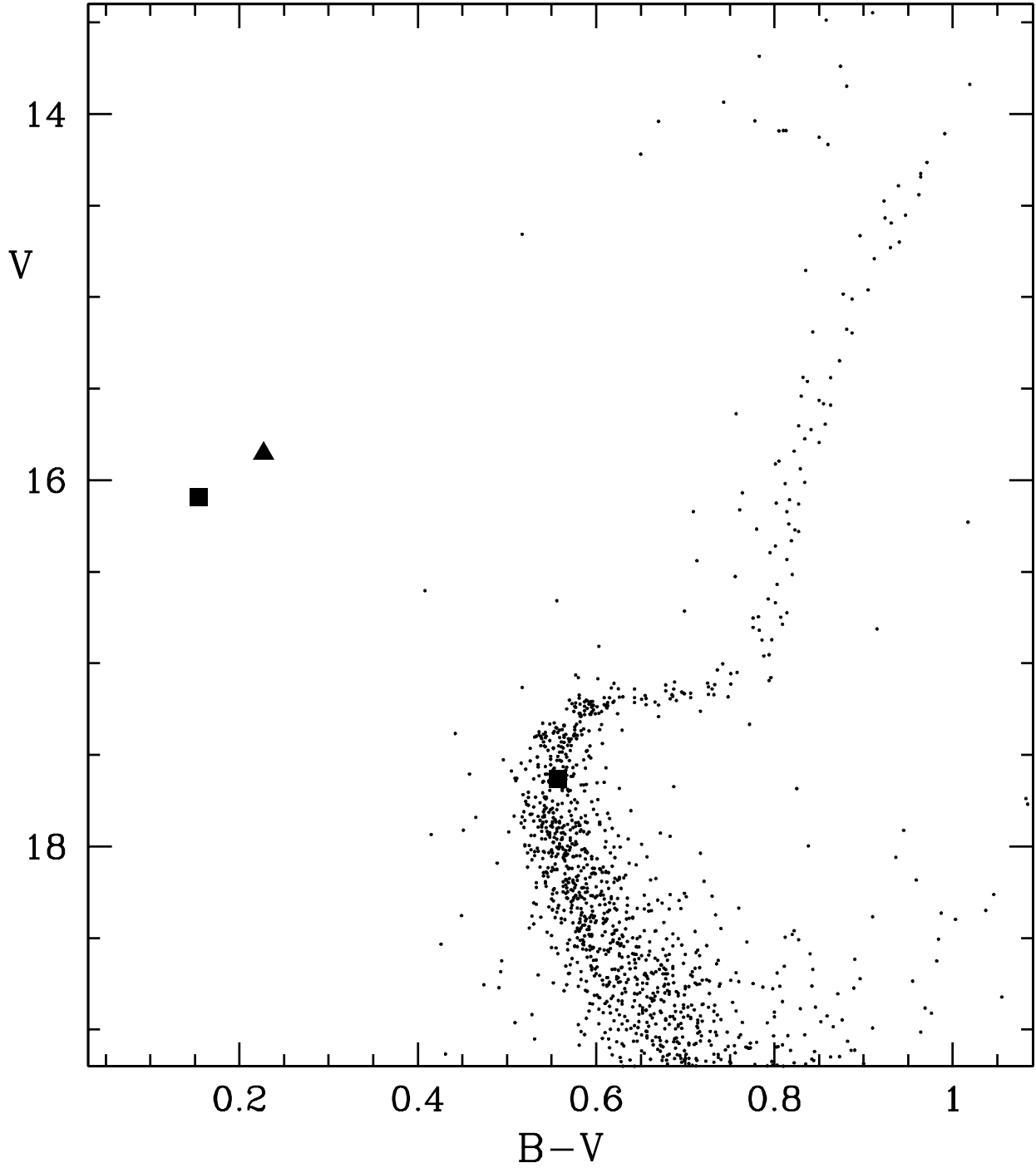}
\caption{
Position of V228 in the color -- magnitude $BV$ diagram for
47~Tuc. The triangle gives the position for the combined
light whereas the squares show the positions of each of the
components separately. 
}
\end{figure}

The absolute parameters of V228 obtained from our spectroscopic and 
photometric analysis are given in Table~\ref{tab6} and the
position of the binary on the color -- magnitude diagram of
47~Tuc is shown in Figure~\ref{fig5}. The uncertainty  of 
temperature $T_1$ includes  estimated uncertainties of both the
photometric zero point of $\delta (B-V) \simeq 0.01$ and of the reddening  
of $\delta E(B-V) \simeq 0.01$. The uncertainty in the
reddening arises from a  comparison
of that commonly used for 47~Tuc,  
$E(B-V)=0.04$ \citep{harris},  with the value of $E(B-V)=0.030$ resulting 
from the  reddening map of \citet{schlegel}.

\section{MEMBERSHIP STATUS}
\label{memb}

Before discussing the evolutionary status of V228
it is worth  considering its membership  in 47~Tuc.
The variable was included in the proper motion study of 47~Tuc
conducted by \citet{tucholke}. According to that study, V228
(designation \#2604) is a genuine proper motion member of the cluster
with probability of 98.2\%.
The systemic velocity of the binary, $V_0 = -22.51$~ km/s
agrees with the radial
velocity of 47~Tuc, $v_{rad}=-18.7\pm 0.5$~km/s  \citep{gebhardt95}.
At the location of the variable  -- about 10 arcmin
from the cluster center -- the velocity dispersion of cluster stars
is about 7~km/s \citep{gebhardt95}.
We conclude that V228 is a radial velocity member of 47~Tuc.
From the light curve solutions one may estimate the observed visual magnitudes
of the components of the binary at maximum light. For the primary
component we obtained $V_1=16.06\pm 0.01$.
Using $M_{\rm V1}=2.66\pm 0.07$ (see Table~\ref{tab6})
one obtains an apparent distance modulus $(m-M)_{\rm V1}=13.40\pm 0.07$
for the primary component of V228. This value is compatible with
several of the recent estimates of the distance of 47~Tuc which
span a range $13.12<(m-M)_{\rm V}<13.55$ \citep{mclaughlin}.
An attempt to obtain a distance modulus for the binary based on the
luminosity of the secondary component is hampered by the difficulty
to  precisely account for ellipsoidal light variations.

In summary, V228 has the same proper motion and radial velocity as 47~Tuc 
and is located at the cluster distance. We conclude that the binary is a 
certain member of 47~Tuc.

\section{EVOLUTIONARY STATUS}
\label{disc}

Our study of the light and radial velocity curves of V228 shows
that the binary belongs to the class of  semi-detached,
low mass, ``cool'' or ``conventional'' Algol variables.
Several well studied systems of this kind are discussed in the literature.
In particular AS~Eri \citep{popper80} and R~CMa \citep{sarma}
are examples of Algols with parameters closely resembling those
of V228. A comprehensive discussion of evolutionary models leading
to the formation of such systems along with a summary of the
observational data is given
by \citet{eggleton} and \citet{nelson}. An earlier but still useful
review of the subject was given by \citet{pacz71}.  According to 
widely accepted  scenarios, the Algol systems form by mass transfer
leading to the reversal of the original mass ratio of the binary so
that the present primaries were originally the less massive components.

The currently secondary component of V228 is noticeably
oversized and overluminous
for its mass. The low value of the observed mass ratio, $q\approx 0.13$, and
the overluminosity of the secondary (exceeding 5 magnitudes) indicate that
the mass transfer occurred in the so-called Case~B 
evolution \citep{pacz71}. In such a case, the original primary
filled its Roche lobe while starting its ascent onto the giant branch.
Its luminosity is currently generated in a hydrogen burning shell
surrounding a degenerate helium core. The present mass transfer in
the binary is expected to occur on a nuclear time scale.
As we have shown above, observations with time base of
11 years failed to reveal any change of the orbital period of V228.

At the first sight the primary component of V228 with $R=1.36\,R_\odot$
seems to be under-sized for its mass of $M=1.51\,M_\odot$.
However, one has to keep in mind that the existing empirical mass-radius
calibrations are based on stars with approximately
solar composition and that stellar models of unevolved stars predict 
a decrease in the radius (for a given mass) for a lower metallicity.
For example, the models of \citet{vandenberg} predict a
ZAMS radius of $R=1.26\,R_\odot$ at $M=1.515\,M_\odot$
and $Z=0.008$ (${\rm [Fe/H]=-0.705}]$) while the models of \citet{girardi}
for the same metallicity 
predict  ZAMS radius of $R=1.33~R_\odot$ for $M=1.5\,M_\odot$.
These models also show that the bolometric
luminosity of the primary is appropriate for an unevolved star
with $M=1.51\,M_\odot$ and the metallicity of 47~Tuc.
In particular, for  $Z=0.008$ and mass $M=1.5\,M_\odot$
\citet{girardi} give $L_{ZAMS}=6.6\,L_\odot$.

The absolute parameters of V228 have implications for the
current turnoff mass of 47~Tuc. According to the evolutionary
model developed by Sarna (in preparation), the binary entered the phase
of mass transfer about 0.2~Gyr ago. Based on the current total mass of
the system of 1.71\,$M_\odot$ we may infer that the original
primary had a mass exceeding $0.85~M_\odot$. This is 
a conservative lower limit assuming  a scenario with perfectly
conservative mass transfer.  Isochrones from 
\citet{vandenberg} for ${\rm [\alpha/Fe]}=+0.3$ and the age of
14~Gyr have turn-off masses of $0.868~M_\odot$ and  $0.852~M_\odot$
for ${\rm [Fe/H]}=-0.606$ and ${\rm [Fe/H]}=-0.707$, respectively.
The observed parameters of V228 together
with an evolutionary interpretation of its current status suggest an upper
limit to the age of 47~Tuc of 14~Gyr. 
This can be compared with a recent age estimate of the cluster
by \citet{gratton}. They obtained an age of 10.8~Gyr using models   
with diffusion and an age of 11.2~Gyr for models with no diffiusion.
If the cluster age is indeed close to 11~Gyr then one has to
conclude that the mass transfer in V228 resulted in a mass
loss from the system.
 
For old stellar systems
like 47~Tuc, it is expected that relatively more massive stars such as binaries
should sink into the core region due to mass segregation.
Apparently this is not the case of V228.
Located at a projected distance $r=588$~arcsec
or 28 core radii from the cluster center, V228  belongs to
the ``external'' sub-population of the cluster
blue stragglers as defined by \citet{ferraro}. 
The observed spatial distribution of the blue straggler 
population in 47~Tuc was studied in detail by \citet{mapelli}.
On the basis of extensive simulations
they concluded that a sizeable fraction of these 
objects is formed in the outer regions of the cluster 
from primordial binaries experiencing mass-transfer, 
induced purely by stellar evolution.
This conclusion is further supported by the recent 
detection of a sub-population of carbon/oxygen depleted 
blue stragglers in 47~Tuc \citep{ferraro06}.
In that context it would be worth  determining the orbital parameters of 
V228 in the cluster. As shown by \citet{mclaughlin}, the 
determination of accurate  proper motions for   stars in 47~Tuc
is possible from HST images with a time base of a
few years. Unfortunately the HST archive does not contain any
images of the V228 field.

In Section~\ref{memb} we  estimated the distance of V228 to check its 
membership in the cluster. One may use the reverse approach and
use the binary to obtain a distance estimate for the cluster. The largest source
of  uncertainty  arises from estimates of the effective temperature 
and bolometric correction from the color index $B-V$. 
The surface brightness method may provide a 
more robust and secure determination of distance to V228 than available from
our data \citep{clausen,ribas}. In particular, recent progress in the 
interferometric techniques  has resulted in substantial improvement of 
precise calibrations of surface brightness in the near-IR
bands \citep{kervella, dibenedetto}. It would be useful to
obtain IR photometry for V228 for an accurate and independent
47~Tuc distance determination.

To summarize, 
we have used photometric and spectroscopic observations of the
blue straggler V228, a member of the globular cluster 47~Tuc,
to derive the masses, radii, and luminosities of the
component stars. The resulting masses indicate that V228 is
a blue straggler which formed through a mass transfer in a close binary
system. We derive an upper limit of 14 Gyr for the turnoff age of
47~Tuc.

\acknowledgments

JK, WP and WK were supported by the grants 1~P03D~001~28  and
76/E-60/SPB/MSN/P-03/DWM35/2005-2007 from the Ministry
of Science and Information Society Technologies, Poland.
IBT was supported by NSF grant AST-0507325.
Support from the Natural Sciences and Engineering Council of Canada
to SMR is acknowledged with gratitude.

The authors would like to thank the referee, Dr.\ Giacomo Beccari,  
for very useful suggestions and comments allowing improvement of the 
paper.

\clearpage

\begin{deluxetable}{lccc}
\tablecolumns{6}
\tablewidth{0pt}
\tabletypesize{\normalsize}
\tablecaption{$BV$ Photometry of V228 at Minima and Quadrature
   \label{tab1}}
\tablehead{
\colhead{Phase}            &
\colhead{$V$}              &
\colhead{$B$}              &
\colhead{$B-V$}            
}
\startdata
Max & 15.854(1) & 16.081(2) & 0.227(2) \\
Min~I & 16.239(1) & 16.484(3) & 0.245(3) \\
Min~II & 15.954(1) & 16.146(2) & 0.192(2) \\
\enddata
\end{deluxetable}


\begin{deluxetable}{rccc}   
\tablecolumns{4}
\tablewidth{0pt}
\tabletypesize{\normalsize}
\tablecaption{Times of Minima and $O-C$ Values for V228
   \label{tab2}}
\tablehead{
\colhead{Cycle}            &
\colhead{$T_0$}            &
\colhead{Error}            &
\colhead{$O-C$}            \\
\colhead{}                 &
\colhead{HJD-2400000}      &
\colhead{}                 &
\colhead{}
}
\startdata
-1589.0 &  49236.3820         & 0.0016 &   -0.0022\\
    0.0 &  51064.8204         & 0.0006 &   -0.0002\\
  279.0 &  51385.8615         & 0.0002 &    0.0001\\
  344.0 &  51460.6564         & 0.0002 &   -0.0002\\
  380.5 &  51502.6554         & 0.0005 &    0.0009\\
  968.0 &  52178.6845         & 0.0005 &   -0.0001\\
 1632.0 &  52942.7401         & 0.0002 &   -0.0001\\
 1631.0 &  52941.5893         & 0.0002 &    0.0001\\
 1939.5 &  53296.5766         & 0.0010 &   -0.0006\\
\enddata
\end{deluxetable}

\begin{deluxetable}{cccrcr}
\tablecolumns{8}
\tablewidth{0pt}   
\tabletypesize{\normalsize}
\tablecaption{Radial Velocities of V228 and Residuals from 
the Adopted Spectroscopic Orbit \label{tab3}}  
\tablehead{
\colhead{HJD-2450000}       &
\colhead{phase}             &
\colhead{RV$_1$}            &
\colhead{$(O-C)_1$}         &
\colhead{RV$_2$}            &
\colhead{$(O-C)_2$}
}
\startdata
3179.9358 & 0.134 &  -44.16&    -1.03&   125.07 &   -7.53\\
2872.8162 & 0.233 &  -49.98&    -0.07&   184.85 &    2.25\\
2872.8401 & 0.254 &  -49.69&     0.38&   188.25 &    4.45\\
2872.8702 & 0.280 &  -47.73&     1.87&   179.64 &   -0.96\\
2946.5904 & 0.346 &  -41.87&     3.38&   144.00 &   -5.70\\
2923.5936 & 0.361 &  -43.81&    -0.13&   138.65 &    0.15\\
3282.6263 & 0.377 &  -43.24&    -1.45&   124.33 &   -0.47\\
2923.6196 & 0.383 &  -42.14&    -1.23&   120.58 &    2.18\\
3282.6451 & 0.393 &  -40.39&    -0.81&   113.47 &    4.87\\
2923.6448 & 0.405 &  -38.60&    -0.70&    97.03 &    0.92\\
3282.6624 & 0.408 &  -38.54&    -1.08&    92.80 &   -0.07\\
2923.6656 & 0.423 &  -34.76&     0.45&    74.25 &   -1.69\\
2944.5689 & 0.589 &   -9.02&    -1.17&  -130.08 &    5.62\\
3183.9315 & 0.607 &   -4.75&     0.69&  -157.91 &   -4.31\\
2944.6061 & 0.622 &   -5.46&    -1.98&  -166.76 &    1.24\\
3281.7577 & 0.622 &   -3.65&    -0.30&  -175.19 &   -6.29\\
3280.6190 & 0.633 &   -3.39&    -1.25&  -184.66 &   -6.96\\
2944.6464 & 0.657 &   -2.18&    -2.60&  -200.99 &   -4.89\\
3280.6808 & 0.686 &    2.52&    -0.37&  -214.24 &   -0.64\\
3182.8927 & 0.704 &    4.10&     0.17&  -220.12 &    0.88\\
3280.7041 & 0.706 &    5.11&     1.08&  -224.38 &   -2.68\\
3182.9159 & 0.724 &    3.70&    -1.00&  -222.97 &    3.43\\
3280.7281 & 0.727 &    4.60&    -0.18&  -228.28 &   -1.28\\
2868.7919 & 0.736 &    4.97&     0.03&  -233.58 &   -5.48\\
3182.9385 & 0.744 &    4.61&    -0.42&  -224.97 &    3.73\\
3280.7535 & 0.749 &    6.07&     1.01&  -228.59 &    0.31\\
2868.8263 & 0.766 &    5.86&     0.93&  -230.37 &   -2.47\\
3280.7787 & 0.771 &    4.08&    -0.73&  -227.37 &   -0.37\\
2868.8604 & 0.795 &    4.12&     0.18&  -224.09 &   -3.39\\
2867.7283 & 0.811 &    3.65&     0.62&  -209.28 &    4.72\\
2927.5675 & 0.814 &    5.88&     3.05&  -208.65 &    3.85\\
2868.8945 & 0.825 &    3.60&     1.50&  -205.68 &    1.42\\
2927.5915 & 0.835 &   -0.28&    -1.46&  -199.58 &    0.82\\
2868.9231 & 0.850 &   -0.52&    -0.36&  -191.24 &   -0.74\\
2867.7738 & 0.851 &    2.43&     2.69&  -190.73 &   -0.93\\
2927.6145 & 0.855 &   -0.04&     0.74&  -188.95 &   -3.05\\
2927.6376 & 0.875 &   -2.74&     0.34&  -164.35 &    4.35\\
2867.8024 & 0.876 &   -2.45&     0.63&  -165.83 &    2.87\\
2927.6607 & 0.895 &   -5.10&     0.59&  -145.26 &    3.84\\
2867.8313 & 0.901 &   -6.67&    -0.28&  -137.99 &    5.81\\
\enddata
\end{deluxetable}

\begin{deluxetable}{lc}
\tablecolumns{2}
\tablewidth{0pt}
\tabletypesize{\normalsize}
\tablecaption{Orbital Parameters for V228
   \label{tab4}}
\tablehead{
\colhead{Parameter}       &
\colhead{Value}
}
\startdata
$P$ (days) & 1.15068618(fixed) \\
$T_{0}$~($HJD-2,450,000$) & 1064.82019(fixed) \\
$e$    & 0.0(fixed) \\
Derived quantities: & \null \\   
$a$~($R_\odot$) & 5.529$\pm$0.024\\
$q$  & 0.1321$\pm$0.0042 \\
$M_1+M_2~(M\odot$) & 1.711$\pm$ 0.022 \\  
$V_0~(km~s^{-1})$ & $-22.51\pm$ 0.40 \\

Other quantities: & \null \\
$\sigma_{1}~(km~s^{-1})$&  1.27 \\
$\sigma_{2}~(km~s^{-1})$&  3.61 \\
\enddata  
\end{deluxetable}

\begin{deluxetable}{lccc}
\tablecolumns{4}
\tablewidth{0pt}
\tabletypesize{\normalsize}  
\tablecaption{Light Curve Solution for V228 \label{tab5}}
\tablehead{
\colhead{Parameter}       &
\colhead{V}               &
\colhead{B}               &
\colhead{Adopted}
}
\startdata
$i$~(deg) & 77.03 $\pm$ 0.11 & 77.01 $\pm$ 0.07 & 77.02 $\pm$ 0.06\\
$\Omega_{1}$ & 4.232 $\pm$ 0.130 & 4.215 $\pm$ 0.058 & 4.2178 $\pm$ 0.053\\
$T_{1}$~(K) & 8075 (fixed) & 8075 (fixed) & 8075 (fixed)\\
$T_{2}$~(K) & 5855 $\pm$ 50 & 5782 $\pm$ 45 & 5814 $\pm$ 33\\
$(L_{1}/L_{2})$ & 4.664  $\pm$ 0.034 & 7.361 $\pm$ 0.047 & \null\\
$r_{1pole}$  & 0.2436 $\pm$ 0.0077& 0.2447 $\pm$ 0.0035 & 0.2445 $\pm$ 0.0032\\
$r_{1point}$ & 0.2466 $\pm$ 0.0076& 0.2477 $\pm$ 0.0036 & 0.2475 $\pm$ 0.0032\\
$r_{1side}$  & 0.2457 $\pm$ 0.0075& 0.2468 $\pm$ 0.0036 & 0.2466 $\pm$ 0.0032\\
$r_{1back}$  & 0.2463 $\pm$ 0.0075& 0.2475 $\pm$ 0.0036 & 0.2473 $\pm$ 0.0032\\
$r_{2pole}$  & 0.2063 &  0.2063 & 0.2063\\
$r_{2point}$ & 0.3053 & 0.3053 & 0.3053\\
$r_{2side}$ & 0.2145 & 0.2145 & 0.2145\\
$r_{2back}$ & 0.2461 & 0.2461 &  0.2461\\
rms~(mag) & 0.0070 & 0.083 & \null \\
\enddata
\end{deluxetable}

\begin{deluxetable}{lr}
\tablecolumns{2}
\tablewidth{0pt}
\tabletypesize{\normalsize}
\tablecaption{Absolute Parameters for V228 \label{tab6}}
\tablehead{
\colhead{Parameter}       &
\colhead{Value}
}
\startdata
$M_{1}$~($M_\odot$)  & 1.512$\pm$0.022 \\
$M_{2}$~($M_\odot$)  & 0.200$\pm$0.007 \\
$R_{1}$~($R_\odot$)  & 1.357$\pm$0.019 \\
$R_{2}$~($R_\odot$)  & 1.238$\pm$0.013 \\
$T_{1}$~(K)            & 8075$\pm$131 \\
$T_{2}$~(K)            & 5814$\pm$ 73 \\
$Lbol_{1}$($L_\odot$)& 7.02 $\pm$ 0.50\\
$Lbol_{2}$($L_\odot$)& 1.57$ \pm$ 0.09\\
$M_{bol1}$~(mag) & 2.62 $\pm$ 0.07 \\
$M_{bol2}$~(mag) & 4.25 $\pm$ 0.06 \\
$M_{\rm V1}$~(mag) & 2.66  $\pm$0.07\\
\enddata
\end{deluxetable}

\clearpage

\end{document}